\documentclass[journal,10pt]{IEEEtran}
\usepackage{amsmath}
\usepackage{graphicx}
\usepackage{indentfirst}
\usepackage{cases}
\usepackage[noadjust]{cite}
\usepackage{filecontents}
\usepackage{tabularx,booktabs, caption}
\usepackage{makecell}
\newcommand\norm[1]{\left\lVert#1\right\rVert}

\newcolumntype{C}{>{\centering\arraybackslash}X} 
\usepackage{lipsum}
\usepackage{fancyhdr}
\usepackage[justification=centering]{caption}
\usepackage{amsmath,amssymb,mathtools,bm,etoolbox}
\usepackage{color}
\usepackage{array}
\usepackage{mathtools}
\usepackage[british]{babel}
\usepackage{csquotes}
\usepackage{nccmath}
\usepackage{gensymb}
\usepackage[shortlabels]{enumitem}
\usepackage[nodisplayskipstretch]{setspace}

\let\emptyset\varnothing
\SetLabelAlign{bibright}{\hss\llap{[#1]}}
\newcounter{mynum}

\usepackage{hyperref}
\hypersetup{
     colorlinks   = true,
     citecolor    = blue,
     linkcolor    = blue,
     urlcolor     = black
}

\allowdisplaybreaks

\usepackage{multirow}
\usepackage{balance}

\usepackage{algorithm}
\usepackage{algpseudocode}

\usepackage{subcaption}

\usepackage{eso-pic}
\begin{document}

\AddToShipoutPictureBG*{%
  \AtPageUpperLeft{%
    \put(\LenToUnit{0.5\paperwidth},\LenToUnit{-1.2cm}){%
      \makebox[0pt]{\parbox{0.9\paperwidth}{\scriptsize\centering
        \copyright~2026 IEEE. Personal use of this material is permitted. Permission from IEEE must be obtained for all other uses, in any current or future media, including reprinting/republishing this material for advertising or promotional purposes, creating new collective works, for resale or redistribution to servers or lists, or reuse of any copyrighted component of this work in other works.}}}}}

    \title{Feedback-Efficient Beam--User Association for Near-Field mmWave Hybrid Beamforming Systems}    
    \author{Thuan Van Le, Ngoc-Thanh Nguyen, Nam Van Dinh, Nguyen Cong Luong,
    Vo Nguyen Quoc Bao,~\IEEEmembership{Senior Member,~IEEE}, Xingwang Li,~\IEEEmembership{Senior Member, IEEE}, and Ngo~Hoang~Tu,~\IEEEmembership{Member,~IEEE}
    \thanks{Thuan Van Le, Ngoc-Thanh Nguyen, and Nam Van Dinh are with the Faculty of Electrical and Electronic Engineering, Phenikaa School of Engineering, Phenikaa University, Hanoi 12116, Vietnam (e-mail: thuan.levan@phenikaa-uni.edu.vn, thanh.nguyenngoc@phenikaa-uni.edu.vn, nam.dinhvan@phenikaa-uni.edu.vn).}
    \thanks{Nguyen Cong Luong is with the Phenikaa School of Computing, Phenikaa University, Hanoi 12116, Vietnam (e-mail: luong.nguyencong@phenikaa-uni.edu.vn).}
    \thanks{Vo Nguyen Quoc Bao and Ngo Hoang Tu are with the Faculty of Information Technology, Van Lang School of Technology, Van Lang University, Ho Chi Minh City 70000, Vietnam (e-mail: bao.vnq@vlu.edu.vn, tu.nh@vlu.edu.vn). \textit{(Corresponding author: Ngo Hoang Tu.)}}
    \thanks{Xingwang Li is with the School of Physics and Electronic Information Engineering, Henan Polytechnic University, Jiaozuo 454003, China (e-mail: lixingwangbupt@gmail.com).}
    }%

\maketitle

\begin{abstract}
Near-field multiuser hybrid beamforming (HBF) relies on joint angle--distance codebooks whose size grows with the array aperture, so that reporting one quality metric per codeword can cost more than full channel state information (CSI) feedback. This letter proposes a feedback-efficient beam--user equipment (UE) association framework. Each UE probes a focusing codebook sampled within the effective beamfocusing Rayleigh distance but reports only its $M$ strongest candidates, and the base station associates UEs and beams with an interference-aware proportional-fair metric that reuses the leakage information carried by this report. Simulations show that $M=3$ suffices: the proposed scheme loses at most $0.7\%$ of the sum spectral efficiency against an optimistic full-metric reporting reference while using $0.8\%$ of its feedback, a $99.11\%$ reduction with respect to full-CSI feedback.

\begin{IEEEkeywords}
Low-overhead feedback, beam--user association, hybrid beamforming, near-field communications, mmWave.
\end{IEEEkeywords}
\end{abstract}


\section{Introduction}\label{Sect:Intro}
Large-scale antenna arrays at millimeter-wave (mmWave) frequencies are expected to play a key role in future wireless systems. As the carrier frequency and the array aperture increase, many user equipments (UEs) fall into the near-field region, where the array response depends jointly on angle and distance~\cite{liu2023near}. Focal-distance discrimination then adds a spatial degree of freedom, so that UEs sharing an angular direction may still be separable, and beam acquisition and beam--UE association must be performed over a joint angle--distance focusing codebook rather than an angular-only one. Hybrid beamforming (HBF), which splits precoding between a phase-shifter analog stage and a reduced-dimensional digital stage, is the architecture of interest here.

Current 5G new radio (NR) systems acquire downlink channel state information (CSI) through codebook-based reporting. In frequency division duplex (FDD) deployments the base station (BS) transmits CSI reference signals and each UE quantizes the measured channel using standardized Type-I or Type-II codebooks, where Type-I provides a low-overhead precoder indication and Type-II combines multiple spatial beams with quantized coefficients at higher feedback cost~\cite{3gpp38214}. Both are angular-domain representations without focal-distance information. Replacing them by a polar-domain codebook \cite{9693928} does not solve the multiuser problem on its own, for two reasons. First, these reports indicate a single-UE precoder, and swapping in polar codewords does not specify how the BS should turn them into a joint UE--beam association under the RF-chain constraint, where co-angular codewords at different focal distances are correlated rather than orthogonal as in a DFT codebook. Second, reporting the quality of every codeword scales with the codebook size and, for the array considered here, exceeds the cost of full-CSI feedback. Near-field beam training \cite{10365224,wei2022codebook} lowers the probing cost of a single UE but leaves the multiuser beam--UE association and its feedback unaddressed. The proposed framework instead lets each UE report a compact set of angle--distance candidates with their quality metrics, which the BS uses to schedule UEs and select analog beams under the radio-frequency (RF) chain constraint. This first-stage report does not build the digital precoder, for which the scheduled UEs subsequently report their reduced-dimensional effective channels.

The contributions of this letter are threefold. First, the focusing codebook adapts the polar sampling bounded by the effective beamfocusing Rayleigh distance (EBRD) of \cite{10988573}, devised there for channel-estimation dictionaries, into a hybrid near/far-field beam-selection codebook whose radial grid, and hence size, follows from the array geometry and a single coherence threshold dimensioned by the measured sum spectral efficiency (SE). Second, each UE probes the codebook but reports only a small candidate set, whose size $M$ is dimensioned from the measured assignment infeasibility and need not equal the number of streams. Third, the proportional-fair (PF) beam--UE association is driven by a sequential metric that exploits, at no extra signaling cost, the leakage terms already carried by the compact report together with the codeword correlations known at the base station. Simulations against full-metric reporting, the interference-blind assignment, angular-only reporting and full-CSI benchmarks show a $99.11\%$ feedback reduction at an SE loss of at most $0.7\%$ with respect to an optimistic full-metric reference.

\section{System Model and Signal Processing}\label{Sect:SystemModel}
We consider an FDD downlink near-field multiuser multiple-input single-output system, where instantaneous downlink CSI cannot generally be inferred directly from uplink channel reciprocity. The BS, equipped with a uniform planar array (UPA) of $N_{\mathrm T}=N_x \times N_y$ antennas ($N_x$ horizontal, $N_y$ vertical) and $N_{\mathrm{RF}}$ RF chains, serves $K \ge N_{\mathrm{RF}}$ single-antenna UEs. Let $N_{\mathrm S}$ denote the number of serving data streams at the BS, assuming $N_{\mathrm S} = N_{\mathrm{RF}}$.\footnote{In general $N_{\mathrm S}=\min(N_{\mathrm{RF}},K)$. We consider $K\ge N_{\mathrm{RF}}$, so the BS selects at most $N_{\mathrm S}=N_{\mathrm{RF}}$ UEs \cite{10296924}.} Let $\mathcal{K}=\{1,\ldots,K\}$ and $\mathcal S = \{1,\ldots, N_{\mathrm S} \}$ denote the sets of UEs and data streams, with $k \in \mathcal K$ and $i \in \mathcal S$ throughout.

An HBF architecture is adopted at the BS, consisting of an analog precoder $\mathbf F_{\mathrm{RF}}\in\mathbb C^{N_{\mathrm T}\times N_{\mathrm{RF}}}$ and a digital baseband precoder $\mathbf F_{\mathrm{BB}} = [{\mathbf{f}}_{\mathrm{BB},1},\ldots,{\mathbf{f}}_{\mathrm{BB},N_{\mathrm S}}]\in\mathbb C^{N_{\mathrm{RF}}\times N_{\mathrm S}}$ \cite{6717211,tu2025semi,tu2026multi}. Accordingly, the transmitted signal is $\mathbf F_{\mathrm{RF}}\mathbf F_{\mathrm{BB}}\mathbf{s}$, where $\mathbf s\in\mathbb C^{N_{\mathrm S} \times 1}$ denotes the vector of transmitted data streams, satisfying $\mathbb E[\mathbf s\mathbf s^H]=\mathbf I_{N_{\mathrm S}}$. The total transmit power at the BS, denoted by $P_t$, is constrained by normalizing $\mathbf F_{\mathrm{BB}}$ such that
\begin{align}\label{eq:Pt_constraint}
    \norm{\mathbf F_{\mathrm{RF}}\mathbf F_{\mathrm{BB}}}_F^2 \le P_t.
\end{align}

Within the Rayleigh distance $R_{\mathrm F}=2D^2/\lambda$ \cite{liu2023near}, where $D=d\sqrt{(N_x-1)^2+(N_y-1)^2}$ is the array aperture, wavefronts are spherical and the propagation distance from each element to the UE is nonuniform.

For the BS UPA, the $(m,n)$-th antenna element is located at $\mathbf p_{m,n}=(\tilde m d,\tilde n d,0)$, where $m=0,\ldots,N_x-1$, $n=0,\ldots,N_y-1$, $\tilde m=m-(N_x-1)/2$, $\tilde n=n-(N_y-1)/2$, and $d=\lambda/2$. A focal point with distance $r$, azimuth $\phi$, and zenith angle $\theta$ is located at \cite[Remark~1]{liu2023near}
\begin{equation}
\mathbf p(\phi,\theta,r)
=
r[\sin\theta\cos\phi,\sin\theta\sin\phi,\cos\theta]^T.
\label{eq:p_coordinate}
\end{equation}
With $r_{m,n}(\phi,\theta,r)=\norm{\mathbf p(\phi,\theta,r)-\mathbf p_{m,n}}_2$ the exact propagation distance \cite[Eq.~(14)]{liu2023near}, the exact near-field UPA response stacks the element phases as
\begin{equation}
\mathbf a^{\mathrm{exact}}(\phi,\!\theta,\!r)\!
=
\!\frac{1}{\sqrt{N_{\mathrm T}}}\operatorname{vec}
\Big(\!
\big\{
e^{-j\frac{2\pi}{\lambda} r_{m,n}(\phi,\theta,r)}
\big\}_{\forall m,n}
\!\Big)\!\in\!\mathbb{C}^{N_{\mathrm T}\!\times\! 1},
\label{eq:exact_nf_steering}
\end{equation}
where $\operatorname{vec}(\cdot)$ is the column-wise vectorization operator. Consequently, the channel between the BS and UE $k$ is modeled as a superposition of $L$ spherical-wave components \cite[Eq.~(11)]{liu2023near}, i.e.,
\begin{equation}
\mathbf h_k
=
\sum\nolimits_{\ell=1}^{L}
\alpha_{k,\ell}\,
\mathbf a^{\mathrm{exact}}(\phi_{k,\ell},\theta_{k,\ell},r_{k,\ell}),
\label{eq:nf_channel}
\end{equation}
where $\alpha_{k,\ell}$ denotes the complex path gain, whose distribution is given in Section~\ref{sec:results}, and $\phi_{k,\ell}$, $\theta_{k,\ell}$, and $r_{k,\ell}$ denote the azimuth angle, zenith angle, and propagation distance of path $\ell$ for UE $k$, respectively.

The received signal at UE $k$ is $y_k=\mathbf h_k^H\mathbf F_{\mathrm{RF}}\mathbf F_{\mathrm{BB}}\mathbf s+z_k$, where $z_k\sim\mathcal{CN}(0,\sigma^2)$ is the additive white Gaussian noise. Accordingly, the post-precoding signal-to-interference-plus-noise ratio (SINR) and the achievable SE of UE $k$ on stream $i$ are given, respectively, by
\begin{align}
\label{eq:SINR}
\mathrm{SINR}_{k,i}&=
\frac{|\mathbf h_k^H\mathbf F_{\mathrm{RF}}\mathbf f_{\mathrm{BB},i}|^2}
{\sum_{i' \in \mathcal S \backslash \{ i\}}|\mathbf h_k^H\mathbf F_{\mathrm{RF}}\mathbf f_{\mathrm{BB},i'}|^2+\sigma^2},\\
\label{eq:Rate_UE_Stream}
{\cal R}_{k,i}&=\log_2(1+\mathrm{SINR}_{k,i}).
\end{align}

For beam design, define $\mu=\sin\theta\cos\phi$ and $\nu=\sin\theta\sin\phi$. Applying the Fresnel approximation \cite[Eq.~(10)]{liu2023near} to $r_{m,n}$, i.e., a second-order expansion about the array center that, unlike the axis-separable UPA form \cite[Eq.~(15)]{liu2023near}, retains the cross-axis term, gives
\begin{align}
{\mathbf a}(\mu,\nu,r)
&=
\operatorname{vec}
\Big(
\big\{
e^{j\frac{2\pi}{\lambda}
\delta_{m,n}(\mu,\nu,r)}
\big\}_{\forall m,n}
\Big)\big/\sqrt{N_{\mathrm T}},
\\
\delta_{m,n}(\mu,\nu,r)
&=
d(\tilde m\mu+\tilde n\nu)
-\frac{d^2}{2r}
\left[
\tilde m^2+\tilde n^2
-(\tilde m\mu+\tilde n\nu)^2
\right],
\label{eq:fresnel_nf_steering}
\end{align}
where $\delta_{m,n}$ approximates $r-r_{m,n}$, and $\mathbf a(\mu,\nu,\infty)$ is the far-field response.

The focusing codebook is built separately in the two domains. Angularly, the direction cosines are taken as the centered discrete Fourier transform (DFT) samples $\mu_p=(2p-N_x-1)/N_x$ and $\nu_q=(2q-N_y-1)/N_y$ that fall inside the angular support of Section~\ref{sec:results}, which matches the sampling to the array resolution and satisfies $\mu_p^2+\nu_q^2<1$, yielding $N_{\mathrm d}$ valid angular directions.
Radially, the sampling follows the beamfocusing capability of the array rather than the Rayleigh distance, adopting the EBRD bound of \cite{10988573}. Defining the codeword coherence $\rho(r,r')=\left|\mathbf a^H(\mu,\nu,r)\,\mathbf a(\mu,\nu,r')\right|^2$, the EBRD $r_{\mathrm E}(\mu,\nu)$ is the largest focal distance whose codeword is still distinguishable from its far-field counterpart, i.e., the solution of $\rho(r_{\mathrm E},\infty)=\rho_0$ for a coherence threshold $\rho_0$, a design parameter of the codebook: $\rho_0=0.5$ recovers the classical $3$-dB effective Rayleigh distance \cite{wei2022codebook}, while larger values accept more correlated codewords as distinct and give a longer $r_{\mathrm E}$ and a denser radial grid. For each direction $(\mu_p,\nu_q)$ the radial grid is generated as follows: (\textit{i}) $r_{\mathrm E}$ is obtained by a one-dimensional search of $\rho(r,\infty)$ over $[r_{\min},R_{\mathrm F}]$, where $r_{\min}=0.62\sqrt{D^3/\lambda}$ is the lower bound of the radiating near field; (\textit{ii}) the first focal distance is $r_1=r_{\min}$; (\textit{iii}) given $r_w$, the next focal distance $r_{w+1}$ is the smallest $r>r_w$ with $\rho(r_w,r)\le\rho_0$, so that consecutive codewords are one beam depth apart; (\textit{iv}) the recursion stops when no such $r$ lies below $r_{\mathrm E}$, and a single far-field codeword $\mathbf a(\mu_p,\nu_q,\infty)$ is appended, since focal distances beyond $r_{\mathrm E}$ are no longer resolvable. The resulting hybrid codebook is expressed as
\begin{equation}
{\cal C}=
\big\{\mathbf a(\mu_p,\nu_q,r_w)\big\}_{r_w\le r_{\mathrm E}(\mu_p,\nu_q)}
\cup
\big\{\mathbf a(\mu_p,\nu_q,\infty)\big\},
\label{eq:codebook}
\end{equation}
whose cardinality $N_{\mathrm b}$ follows from the array geometry instead of being chosen by hand. Codebooks uniform in $r$, uniform in $1/r$ \cite{9693928}, and angular-only DFT codebooks are kept as baselines in Section~\ref{sec:results}.

At each transmission time interval (TTI), the BS can form the analog precoder $\mathbf F_{\mathrm{RF}}$ after solving the scheduling problem, formulated in Section~\ref{sec:section3}. Let $\mathcal U = \{ u_1,\ldots, u_{N_{\mathrm S}} \}$ denote the set of scheduled UEs, where $u_i = k$ indicates that stream $i$ is assigned to UE $k$. Given $\mathcal U$, the corresponding effective channel matrix is expressed as
\begin{equation}
\label{eq:H_eff}
\mathbf H_{\mathrm{eff}}=
\big[\mathbf h_{u_1},\ldots,\mathbf h_{u_{N_{\mathrm S}}}\big]^H\mathbf F_{\mathrm{RF}} \in \mathbb{C}^{N_{\mathrm S} \times N_{\mathrm{RF}}}.
\end{equation}
The digital precoder is then designed using regularized zero-forcing (RZF) as
\begin{equation}
\mathbf F_{\mathrm{BB}}=
\beta\mathbf H_{\mathrm{eff}}^H
\left(\mathbf H_{\mathrm{eff}}\mathbf H_{\mathrm{eff}}^H+\xi\mathbf I\right)^{-1},
\label{eq:cal_FBB}
\end{equation}
where $\xi=N_{\mathrm S}\sigma^2/P_t$ is the RZF regularization term and $\beta = \sqrt{ {P_t}\Big/{ \norm{ \mathbf{F}_{\mathrm{RF}} \mathbf{H}_{\mathrm{eff}}^{H} \left( \mathbf{H}_{\mathrm{eff}}\mathbf{H}_{\mathrm{eff}}^{H} + \xi \mathbf{I} \right)^{-1} }_F^2 } }$ is the normalization factor chosen to satisfy \eqref{eq:Pt_constraint}.

The compact report of Section~\ref{sec:section3} serves only scheduling and analog beam selection. Once $\mathbf F_{\mathrm{RF}}$ is fixed, each scheduled UE estimates $\mathbf g_k^{H}=\mathbf h_k^{H}\mathbf F_{\mathrm{RF}}$ from dedicated pilots and reports its quantized coefficients, which the BS stacks into $\mathbf H_{\mathrm{eff}}$.

\section{Proposed Strategy}\label{sec:section3}
Since the near-field focusing codebook $\mathcal C=\{\mathbf c_1,\ldots,\mathbf c_{N_{\mathrm b}}\}$, indexed by $\mathcal N=\{1,\ldots,N_{\mathrm b}\}$, is known to both the BS and UEs, each UE exhaustively probes all $N_{\mathrm b}$ codewords and locally evaluates their beam qualities. On the pilot transmitted through codeword $b$ with power $P_t/N_{\mathrm S}$, UE $k$ measures the complex gain $\hat g_{k,b}$, an estimate of $\mathbf h_k^H\mathbf c_b$ that requires no knowledge of $\mathbf h_k$ itself, and forms the signal-to-noise ratio (SNR)-like ranking metric as
\begin{align}
\label{eq:Gamma_kb}
\Gamma_{k,b}
&=
{P_t|\hat g_{k,b}|^2}\big/{(N_{\mathrm S}\sigma^2)}.
\end{align}
The metric serves ranking only and is not the post-precoding SINR. Rather than feeding back only the maximum-SNR codeword, UE $k$ selects its $M$ strongest joint angle--distance candidate beams, denoted by $\mathcal M_k=\{b_k^{(1)},\ldots,b_k^{(M)}\}$, the indices of the $M$ largest values in $\{\Gamma_{k,b}\}_{b\in\mathcal N}$, ordered such that $\Gamma_{k,b_k^{(1)}}\ge \cdots \ge \Gamma_{k,b_k^{(M)}}$. UE $k$ then feeds back the $M$ pairs $(b,\Gamma_{k,b})$, $b\in\mathcal M_k$, to the BS. The report size $M$ is a design parameter that need not equal $N_{\mathrm S}$, and it is dimensioned from the measured assignment infeasibility in Section~\ref{sec:results}.

After collecting the reports, the BS incorporates PF scheduling to improve long-term fairness. The PF metric for the candidate beam $b\in\mathcal M_k$ at UE $k$ is defined as \cite{10296924}
\begin{equation}
\Lambda_{k,b}={\log_2(1+\Gamma_{k,b})}\big/{\bar{\cal R}_{k}},
\label{eq:pfmetric_coml2}
\end{equation}
where ${\bar{\cal R}_{k}}$ denotes the recursively updated rate of UE $k$ across TTIs, with the update rule \cite{10296924}
\begin{align}
\bar{\cal R}_{k}(t)
=
(1-\eta) \bar{\cal R}_{k}(t-1)+\eta {\cal R}_{k}(t-1).
\label{eq:ravg_update_coml2}
\end{align}
Here, $t$ is the TTI index, so that $\bar{\cal R}_k(t)$ is the average available when the association of TTI $t$ is computed, $\eta\in(0,1)$ is the PF averaging factor, and $\mathcal R_k(t-1)$ is the rate ${\cal R}_{k,i}$ in \eqref{eq:Rate_UE_Stream} achieved by UE $k$ on its assigned stream $i$ at TTI $t-1$, set to zero for unscheduled UEs.

The BS maximizes the sum PF metric under the one-UE-per-stream constraint as follows:
\begin{subequations}
\label{eq:OptimizationProblem}
\begin{align}
\label{eq:OptimizationProblem_a}
\max_{\mathcal X}\quad &\sum\nolimits_{k\in \mathcal K}\sum\nolimits_{b \in {\mathcal M_k}}x_{k,b}\Lambda_{k,b}\\
\label{eq:OptimizationProblem_b}
\mathrm{s.t.}\quad
&\sum\nolimits_{b \in {\mathcal M_k}}x_{k,b}\le 1,\ \forall k  \in {\cal K},\\
\label{eq:OptimizationProblem_c}
&\sum\nolimits_{k\in\mathcal K} x_{k,b} \le 1, \quad \forall b\in\textstyle\bigcup_{k'\in\mathcal K}\mathcal M_{k'},\\
\label{eq:OptimizationProblem_d}
&\sum\nolimits_{k\in \cal K}\sum\nolimits_{b \in {\mathcal M_k}}x_{k,b} = N_{\text{RF}},\\
&
\label{eq:OptimizationProblem_e}
x_{k,b}\in\{0,1\},
\end{align}
\end{subequations}
where ${\cal X} = \big\{ {{{\left( {{x_{k,b}}} \right)}_{\forall k \in {\cal K},b \in {\mathcal M_k}}}} \big\}$ denotes the set of binary optimization variables, with $x_{k,b}=1$ indicating that beam $b$ is assigned to UE $k$, and $x_{k,b}=0$ otherwise. Constraints \eqref{eq:OptimizationProblem_b} and \eqref{eq:OptimizationProblem_c} assign each UE to at most one beam and each beam to at most one UE, \eqref{eq:OptimizationProblem_d} enforces exactly $N_{\mathrm S}=N_{\text{RF}}$ active pairs, and \eqref{eq:OptimizationProblem_e} imposes binary variables.

\begin{algorithm}[t!]
\small
\caption{\small Proposed Compact Feedback and Interference-Aware Beam--UE Association}
\label{Alg:Proposed}
\begin{algorithmic}[1]

\Statex \hspace{-0.6cm} \textit{\textbf{UE-Side Beam Probing and Low-Overhead Feedback}}
\For{each UE $k\in\mathcal K$}
    \State Evaluate $\Gamma_{k,b}$, $\forall b \in \mathcal N$, by \eqref{eq:Gamma_kb} and feed back $(b,\Gamma_{k,b})$, $b\in\mathcal M_k$;
\EndFor

\Statex \hspace{-0.6cm} \textit{\textbf{BS-Side Sequential Association and HBF}}
\State Initialize $\mathcal A\leftarrow\emptyset$, $\mathcal B\leftarrow\emptyset$, and the unassigned-UE set $\mathcal K'\leftarrow\mathcal K$;
\For{$i=1,\ldots,N_{\mathrm{RF}}$}
    \State Form the candidate set $\mathcal P(\mathcal A)=\{(k,b):k\in\mathcal K',\ b\in\mathcal M_k\setminus\mathcal B\}$;
    \State \textbf{if} $\mathcal P(\mathcal A)=\emptyset$ \textbf{then} declare the report infeasible and \textbf{break};
    \State $(k^\star,b^\star)=\arg\max_{(k,b)\in\mathcal P(\mathcal A)}\Psi_{k,b}(\mathcal A)$ by \eqref{eq:assoc_metric};
    \State $\mathcal A\leftarrow\mathcal A\cup\{(k^\star,b^\star)\}$, $\mathcal B\leftarrow\mathcal B\cup\{b^\star\}$, $\mathcal K'\leftarrow\mathcal K'\setminus\{k^\star\}$;
\EndFor
\State \textbf{while} $|\mathcal A|<N_{\mathrm{RF}}$: append the UE of smallest $\bar{\cal R}_k$ in $\mathcal K'$ with the unselected codeword outside its report least correlated with $\mathcal B$;
\State Set $x_{k,b}=1$ for $(k,b)\in\mathcal A$ and $\mathbf F_{\mathrm{RF}}=[\mathbf c_b]_{(k,b)\in\mathcal A}$;
\State Acquire the quantized effective channels of the scheduled UEs, build $\mathbf H_{\mathrm{eff}}$ by \eqref{eq:H_eff}, compute $\mathbf F_{\mathrm{BB}}$ by \eqref{eq:cal_FBB}, and \Return $\mathbf F_{\mathrm{RF}}$, $\mathbf F_{\mathrm{BB}}$.
\end{algorithmic}
\end{algorithm}

The PF metric \eqref{eq:pfmetric_coml2} scores each candidate pair in isolation, so \eqref{eq:OptimizationProblem} may populate $\mathbf F_{\mathrm{RF}}$ with strongly interfering or near-duplicate codewords, which leaves $\mathbf H_{\mathrm{eff}}$ ill-conditioned for RZF. The proposed association therefore scores each candidate against the pairs already selected. For a partial assignment $\mathcal A$ with selected beam set $\mathcal B$, the utility of assigning beam $b\in\mathcal M_k$ to UE $k$ is defined as
\begin{align}
\Psi_{k,b}(\mathcal A)
&=
\log_2\!\Big(1+\frac{\Gamma_{k,b}}{1+\sum\nolimits_{b'\in\mathcal B\cap\mathcal M_k}\Gamma_{k,b'}}\Big)/\bar{\cal R}_k\nonumber\\
&\quad\times \Big(1-\max\nolimits_{b'\in\mathcal B}\big|\mathbf c_b^H\mathbf c_{b'}\big|^2\Big).
\label{eq:assoc_metric}
\end{align}
The first factor derates the PF metric by the leakage that UE $k$ itself reported for the beams already in use, so the extra entries of the compact report act as a coarse beam-domain interference profile at no extra signaling cost. The second discourages codewords strongly correlated with those already selected and uses only the codebook known at the BS. Both derating terms vanish when $\mathcal A=\emptyset$, the maximum over $\emptyset$ being zero, so \eqref{eq:assoc_metric} then reduces to \eqref{eq:pfmetric_coml2}. The near-field structure enters through the second factor: two codewords sharing a direction but differing in focal distance satisfy $|\mathbf c_b^H\mathbf c_{b'}|^2\approx1$ in the far field, whereas within the EBRD their coherence is at most $\rho_0$ for adjacent focal distances by step (\textit{iii}) and below $0.23$ for the others in the adopted codebook, so co-angular UEs can be multiplexed by focal distance. Because $\Psi_{k,b}$ depends on the selected pairs, \eqref{eq:OptimizationProblem} with $\Lambda_{k,b}$ replaced by $\Psi_{k,b}(\mathcal A)$ is no longer a linear assignment and the bipartite solver of \cite{10296924} does not apply. Steps~4--10 of Algorithm~\ref{Alg:Proposed} approximate it greedily. Starting from $\mathcal A=\emptyset$, each of the $N_{\mathrm{RF}}$ iterations appends the reported pair of largest $\Psi_{k,b}(\mathcal A)$ among the UEs and beams not yet assigned, which enforces \eqref{eq:OptimizationProblem_b}--\eqref{eq:OptimizationProblem_c} by construction, and the loop ends after $N_{\mathrm{RF}}$ appends, which enforces \eqref{eq:OptimizationProblem_d}. Each iteration evaluates at most $KM$ utilities, with both derating terms updated incrementally from the precomputed codeword correlations, so the cost is ${\cal O}(N_{\mathrm{RF}}KM)$, independent of $N_{\mathrm b}$. If the candidate set empties before $N_{\mathrm{RF}}$ pairs are found, the reported graph is infeasible and Step~11 fills the vacant chains with codewords outside the report, hence outside \eqref{eq:OptimizationProblem}. The probability of this event dimensions $M$. Removing the leakage sum and the correlation term from \eqref{eq:assoc_metric} gives $\Lambda_{k,b}$, and solving \eqref{eq:OptimizationProblem} with $\Lambda_{k,b}$ optimally is the interference-blind bipartite assignment of \cite{10296924}, denoted PF-BDP and retained as a benchmark.

Acquisition has three components. Probing the $N_{\mathrm b}$ codewords is beam-training overhead, not reduced by the proposed method. All UEs then report their $M$ candidate-beam indices and quality metrics for scheduling and analog beam selection, and finally the $N_{\mathrm S}$ scheduled UEs report their $N_{\mathrm{RF}}$-dimensional effective channels for digital precoding. Hence, the total feedback is given by
\begin{equation}
O_{\mathrm{Proposed}}
=
KM\!\left(\left\lceil\log_2N_{\mathrm b}\right\rceil+Q_\Gamma\right)
+
N_SN_{\mathrm{RF}}Q_{\mathrm{eff}},
\label{eq:feedback_overhead}
\end{equation}
where $Q_\Gamma$ is the number of bits per reported quality metric, $Q_{\mathrm A}$ and $Q_{\mathrm P}$ are those for the amplitude and phase of an effective-channel coefficient, and $Q_{\mathrm{eff}}=Q_{\mathrm A}+Q_{\mathrm P}$. Full-CSI feedback instead costs $KN_{\mathrm T}Q_{\mathrm{Full}}$ bits, with $Q_{\mathrm{Full}}$ bits per channel coefficient. The ranking metric \eqref{eq:Gamma_kb} and the greedy approximation of \eqref{eq:OptimizationProblem} are not jointly optimal, but they give a practical low-complexity design for large focusing codebooks.

\section{Simulation Results}\label{sec:results}
With $N_x\times N_y=128\times8$, the Rayleigh distance is $R_{\mathrm F}=80.9$~m, and every UE is dropped inside $[r_{\min},R_{\mathrm F}]$ with $r_{\min}=3.14$~m, so that the whole service area is in the near field. A fraction $\zeta=0.75$ of the UEs lies in $N_{\mathrm h}=3$ co-angular clusters, with Gaussian direction-cosine jitter of $0.1$ beamwidths ($2/N_x$ in $\mu$, $2/N_y$ in $\nu$) and ranges drawn inside the local EBRD. The rest are uniform over $\mu\in[-\sin60^\circ,\sin60^\circ]$, $\nu\in[-\sin30^\circ,\sin30^\circ]$, $r\in[r_{\min},R_{\mathrm F}]$. Clusters are included because co-angular UEs are the case in which range focusing, rather than angular steering, provides the separability. The dominant path has $|\alpha_{k,1}|^2=\kappa/(\kappa+1)$ with Rician factor $\kappa=10$~dB and uniform phase, the other $L-1$ paths have $\alpha_{k,\ell}\sim\mathcal{CN}(0,1/((\kappa+1)(L-1)))$ and Gaussian offsets of $0.5$ beamwidths in direction cosine and $5\%$ in range, and each channel is normalized to $\|\mathbf h_k\|=1$, i.e., path loss is not modeled. The transmit SNR $P_t/\sigma^2$ is thus the SNR that a UE attains with an ideal matched beam. UE positions and channels are redrawn in every TTI, $\bar{\cal R}_k$ is initialized to $10^{-3}$, and $\Gamma_{k,b}$ is quantized uniformly in dB over $[-30,30]$~dB. Channel generation uses the exact spherical response \eqref{eq:exact_nf_steering}, whereas the codebook uses the Fresnel expansion \eqref{eq:fresnel_nf_steering}, so the results include the model mismatch, and the probing gains $\hat g_{k,b}$ are taken error-free, as with sufficient pilot averaging, before their $Q_\Gamma$-bit quantization. The remaining parameters are listed in Table~\ref{tab:parameters}.

\begin{table}[!t]
\footnotesize
\centering
\caption{Simulation Parameters.}
\label{tab:parameters}
\begin{tabular}{l c}
\hline
\textbf{Parameter} & \textbf{Value} \\ \hline
Transmit SNR $P_t/\sigma^2$ & 6 dB\\
$f_c$, $N_x\times N_y$, $K$, $N_{\mathrm{RF}}$, $L$ & 30~GHz, $128\times8$, 16, 8, 3 \\
$R_{\mathrm F}$; range support & $80.9$~m; $[3.14,\,80.9]$~m \\
$\rho_0$; $N_{\mathrm d}$; $N_{\mathrm b}$; $M$ & $0.7$; 440; 1904; 3 \\
$(Q_\Gamma,Q_{\mathrm A},Q_{\mathrm P},Q_{\mathrm{Full}})$ & $(6,5,5,10)$~bits \\
$(\zeta,N_{\mathrm h})$; $\eta$; TTIs & $(0.75,3)$; $0.01$; 2000 \\
\hline
\end{tabular}
\end{table}

The proposed method is compared with six benchmarks. The first three share the probing metric \eqref{eq:Gamma_kb} and the RZF precoder \eqref{eq:cal_FBB} with it, so that only the reporting stage or the beam set differs. \textbf{FULL-FB} reports all $N_{\mathrm b}$ metrics and runs the identical association rule, retaining in every slot the full-report or compact solution of higher instantaneous sum-SE, so it is an explicitly optimistic complete-feedback reference. \textbf{PF-BDP} applies the interference-blind bipartite assignment of \cite{10296924} to the same top-$M$ report, isolating the contribution of \eqref{eq:assoc_metric}, while \textbf{DFT-FB} keeps the pipeline but replaces the focusing codebook by an angular-only DFT codebook, isolating the contribution of range focusing. Three benchmarks use quantized full-dimensional CSI and act as feedback-unconstrained references: \textbf{SUS} applies semi-orthogonal user selection \cite{1603708} and gives each selected UE its strongest free focusing codeword, \textbf{FC-OMP} selects analog beams by orthogonal matching pursuit over the focusing codebook \cite{6717211} and assigns UEs by PF, and \textbf{FC-FDZ} applies fully digital zero-forcing-type (RZF) precoding \cite{spencer2004zero} to the SUS-selected UEs.

\begin{figure}[t]
\centering
\includegraphics[width=.85\linewidth]{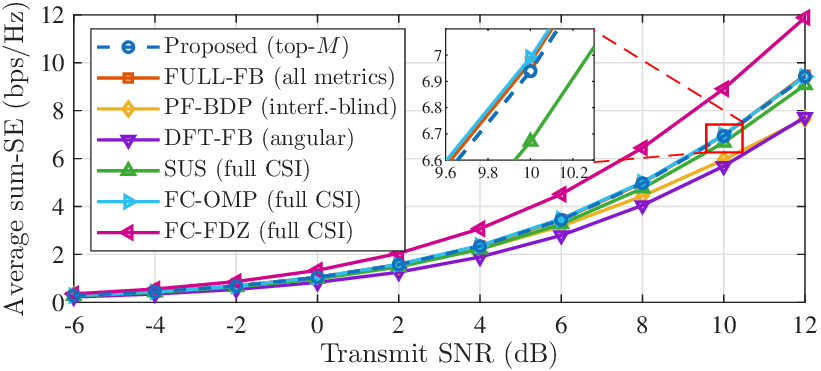}
\caption{Average sum-SE versus transmit SNR.}
\label{fig:throughput}
\end{figure}

Fig.~\ref{fig:throughput} shows the average sum-SE versus the transmit SNR, with SE gaps expressed relative to the proposed scheme. The proposed scheme stays within $0.6\%$ of FULL-FB over the entire range, even though FULL-FB is granted the better of the two solutions in every slot, since the discarded metrics belong to codewords weaker than each UE's $M$-th strongest. The gap to PF-BDP widens from $7.7\%$ at $6$~dB to $18.7\%$ at $12$~dB, because the interference-blind metric leaves $\mathbf H_{\mathrm{eff}}$ ill-conditioned once the system is interference-limited. For SNR $\le-2$~dB, PF-BDP exceeds the proposed scheme by up to $2.3\%$, as the leakage derating penalizes strong codewords in the noise-limited regime. The gap to DFT-FB lies between $18.0\%$ and $21.2\%$, since an angular-only codebook cannot multiplex co-angular UEs. Among the full-CSI references, SUS lies $5.0\%$ below the proposed scheme at $6$~dB, FC-OMP is only $1.2\%$ above it at $6$~dB and is matched at $12$~dB, while FC-FDZ is the highest throughout at the cost of $N_{\mathrm T}$ RF chains and full-dimensional feedback.

\begin{figure}[t]
\centering
\includegraphics[width=.85\linewidth]{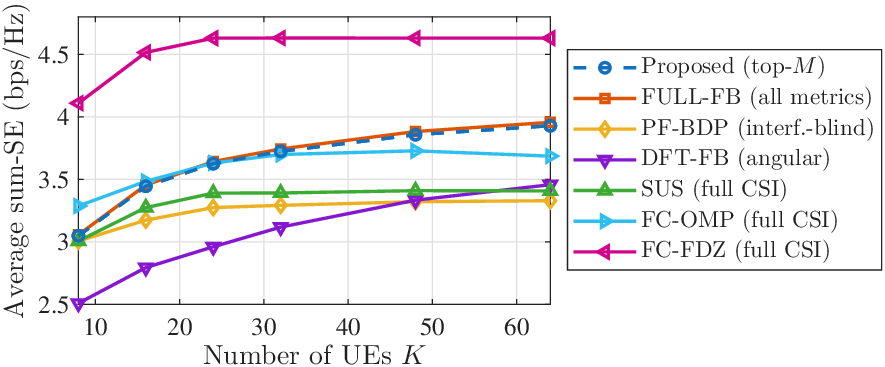}
\caption{Average sum-SE versus the number of UEs.}
\label{fig:throughputK}
\end{figure}

Fig.~\ref{fig:throughputK} shows the average sum-SE versus the number of UEs $K$. All schemes gain from multiuser diversity up to $K=24$. Beyond it, PF-BDP, SUS and FC-FDZ saturate and FC-OMP peaks at $K=48$, whereas the proposed scheme and DFT-FB keep improving. The proposed scheme tracks FULL-FB to within $0.7\%$, while the gap to PF-BDP widens from $1.4\%$ at $K=8$ to $15.3\%$ at $K=64$, because a larger candidate pool lets \eqref{eq:assoc_metric} pick pairs whose codewords are mutually less correlated. DFT-FB remains $12.0\%$--$18.9\%$ below throughout. At $K=64$, SUS lies $13.3\%$ below the proposed scheme. The proposed scheme overtakes FC-OMP for $K\ge32$, and FC-OMP lies $6.2\%$ below it at $K=64$. The leakage derating does not cost PF fairness: over windows of $1/\eta=100$ TTIs, Jain's index of the UE rates stays above $0.994$ for the proposed scheme and $0.995$ for PF-BDP in both sweeps, and the lowest windowed UE rate of PF-BDP never exceeds that of the proposed scheme, lying $9.1\%$ below it at $6$~dB.

\begin{figure}[t]
\centering
\includegraphics[width=.85\linewidth]{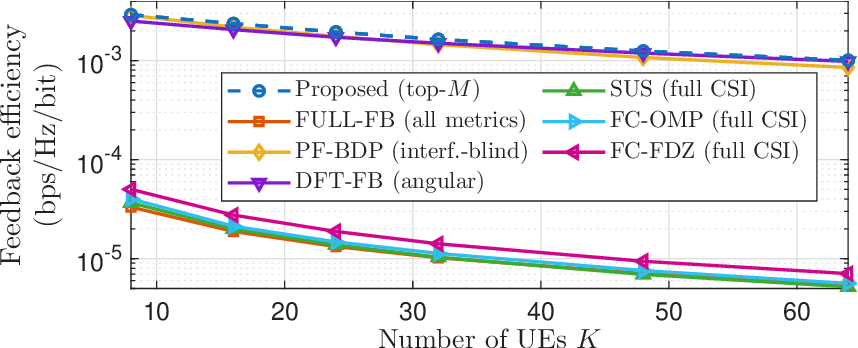}
\caption{Feedback efficiency versus the number of UEs.}
\label{fig:feedback_efficiency}
\end{figure}

Fig.~\ref{fig:feedback_efficiency} shows the feedback efficiency, defined as the ratio of average sum-SE to the total number of feedback bits. The proposed method achieves the highest feedback efficiency over the entire range of $K$, since each UE reports only $M$ index--metric pairs. Within Table~\ref{tab:parameters}, the proposed scheme needs $16\cdot3\cdot(11+6)+8\cdot8\cdot10=1456$~bits/TTI, against $16\cdot1024\cdot10=163840$ for full-CSI feedback and $16\cdot1904\cdot6+640=183424$ for FULL-FB, i.e., reductions of $99.11\%$ and $99.21\%$. These counts exclude probing, a broadcast cost common to all UEs, which uses $N_{\mathrm b}=1904$ pilot beams against $N_{\mathrm T}=1024$ ports for full CSI.

\begin{figure}[t]
\centering
\begin{subfigure}{0.32\linewidth}
    \centering
    \includegraphics[width=\linewidth]{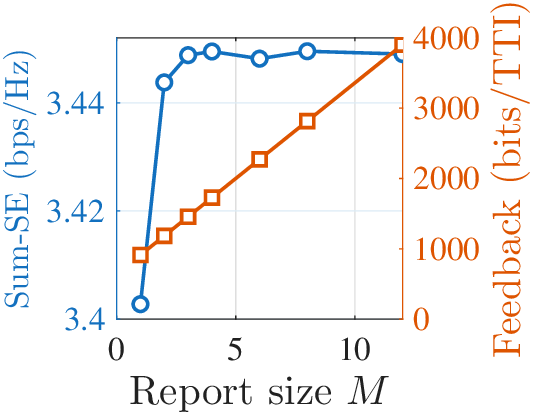}
    \caption{}
\end{subfigure}
\begin{subfigure}{0.32\linewidth}
    \centering
    \includegraphics[width=\linewidth]{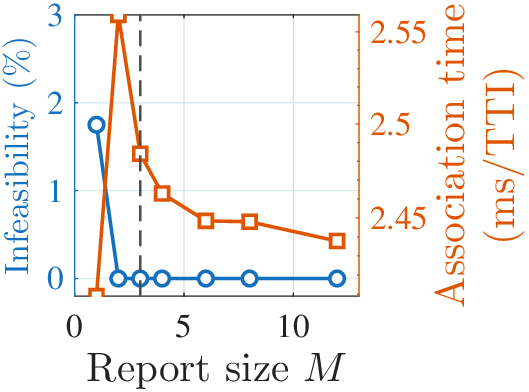}
    \caption{}
\end{subfigure}
\begin{subfigure}{0.32\linewidth}
    \centering
    \includegraphics[width=\linewidth]{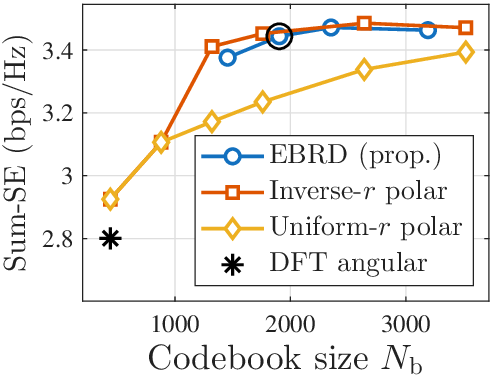}
    \caption{}
\end{subfigure}
\caption{Design study: (a) sum-SE and feedback versus $M$; (b) infeasibility and association time versus $M$; (c) sum-SE versus codebook size $N_{\mathrm b}$ per family.}
\label{fig:throughput_ant_conf}
\end{figure}

Fig.~\ref{fig:throughput_ant_conf} reports the design study. Fig.~\ref{fig:throughput_ant_conf}(a) shows that the sum-SE rises from $3.403$ to $3.449$~bps/Hz as $M$ grows from $1$ to $3$ and is flat afterwards, within $0.02\%$. The additional reported entries feed the leakage term of \eqref{eq:assoc_metric}, and three candidates already carry the useful interference information, whereas the feedback grows linearly from $912$ ($M=1$) to $3904$~bits/TTI ($M=12$). Fig.~\ref{fig:throughput_ant_conf}(b) shows that the assignment infeasibility equals $1.75\%$ at $M=1$ and vanishes for $M\ge2$, while the measured association time stays within $2.4$--$2.6$~ms/TTI for every $M$, since our MATLAB implementation evaluates \eqref{eq:assoc_metric} over the full $K\times N_{\mathrm b}$ metric matrix with unreported entries masked. We therefore adopt $M=3$, the smallest value at which the sum-SE has saturated, one step above the feasibility threshold. Fig.~\ref{fig:throughput_ant_conf}(c) compares the codebook families as curves against $N_{\mathrm b}$. Within the proposed family, the sum-SE is flat to
within $0.8\%$ over $\rho_0\in[0.7,0.9]$, and we adopt $\rho_0=0.7$, the smallest codebook in the flat region. The classical value $\rho_0=0.5$ lies below it, losing $2.0\%$ of the sum-SE. The inverse-$r$ family of \cite{9693928} performs on par, lying $0.2\%$ above the adopted point with $N_r=4$ focal distances per direction ($N_{\mathrm b}=1760$) and $1.2\%$ above it at its best ($N_r=6$, $N_{\mathrm b}=2640$). The EBRD construction reaches this level with a direction-dependent number of focal distances, from $1$ to $5$ here, set by $\rho_0$ instead of a common $N_r$. The best uniform-$r$ codebook ($N_r=8$, $N_{\mathrm b}=3520$) lies $1.5\%$ below the adopted point, which needs $45.9\%$ fewer codewords to probe. The DFT codebook lies $18.7\%$ below the operating point. Sizing the radial grid from the EBRD rather than uniformly up to $R_{\mathrm F}$ therefore saves probing overhead.

\section{Conclusion}
This letter proposed a low-overhead feedback and beam--UE association framework for near-field multiuser mmWave HBF, combining an EBRD-aware focusing codebook, a compact top-$M$ report, and an interference-aware PF association. Simulation results revealed that a report of $M=3$ candidates suffices: the sum-SE stays within $0.7\%$ of an optimistic full-metric reporting reference while the total feedback falls by $99.21\%$ with respect to it and by $99.11\%$ with respect to full-CSI feedback. Replacing the proposed association metric by its interference-blind counterpart costs up to $18.7\%$ of the sum-SE, and an angular-only codebook costs up to $21.2\%$. Sizing the radial grid from the EBRD performs on par with the inverse-$r$ design and exceeds the best uniformly sampled polar codebook by $1.5\%$ with $45.9\%$ fewer codewords to probe.

\bibliographystyle{IEEEtran}
\bibliography{reference}
\end{document}